\documentclass[reprint,superscriptaddress,showpacs,amsmath,amssymb,aps,prb]{revtex4-1}

\usepackage{graphicx}
\usepackage{dcolumn}
\usepackage{bm}
\usepackage{hyperref}

\begin{document}


\title{Evidence of a full gap in LaFeAsO$_{1-x}$F$_x$ thin films from infrared spectroscopy}

\author{Xiaoxiang Xi}
\affiliation{Photon Sciences, Brookhaven National Laboratory, Upton, NY 11973, USA}
\author{Y. M. Dai}
\affiliation{Condensed Matter Physics and Materials Science Department, Brookhaven National Laboratory, Upton, New York 11973, USA}
\author{C. C. Homes}
\affiliation{Condensed Matter Physics and Materials Science Department, Brookhaven National Laboratory, Upton, New York 11973, USA}
\author{M. Kidszun}
\author{S. Haindl}
\affiliation{Institute for Solid State Research, Leibniz Institute for Solid State and Materials Research Dresden, Helmholtzstr. 20, 01069 Dresden, Germany}
\author{G. L. Carr}
\affiliation{Photon Sciences, Brookhaven National Laboratory, Upton, NY 11973, USA}
\date{\today}

\begin{abstract}
We report conventional and time-resolved infrared spectroscopy on LaFeAsO$_{1-x}$F$_x$ superconducting thin films. The far-infrared transmission can be quantitatively explained by a two-component model including a conventional s-wave superconducting term and a Drude term, suggesting at least one carrier system has a full superconducting gap. Photo-induced studies of excess quasiparticle dynamics reveal a nanosecond effective recombination time and temperature dependence that agree with a recombination bottleneck in the presence of a full gap. The two experiments provide consistent evidence of a full, nodeless though not necessarily isotropic, gap for at least one carrier system in LaFeAsO$_{1-x}$F$_x$.
\end{abstract}
\maketitle

Iron-based superconductors have complex gap symmetry and structure that have been under intense study.\cite{Johnston2010} Establishing a unified picture holds the key to understanding the superconductivity mechanism, and this remains a challenge. Among the various iron-based superconductors, LaFeAsO$_{1-x}$F$_x$ is the first found with a $T_c$ higher than most conventional superconductors having phonon-mediated pairing.\cite{Kamihara2008} However it is not the one being most thoroughly studied due to the difficulty in synthesizing high-quality single crystals.\cite{Karpinski2009} The maximum $T_c$ observed in this oxypnictide family ($>50$~K\cite{Wen2011}) exceeds that expected for a phonon-mediated system, indicating unconventional pairing.

In LaFeAsO$_{1-x}$F$_x$, superconductivity arises when fluorine doping suppresses the antiferromagnetic spin-density-wave state and the structural distortion of the parent compound.\cite{Clarina2008} Density functional theory calculations\cite{Singh2008} and ARPES\cite{Lu2009} have established a multiband electronic structure in LaFeAsO, with hole pockets at the $\Gamma$ point of the Brillouin zone and electron pockets at the M points, suggesting the possibility for multiple superconducting gaps in LaFeAsO$_{1-x}$F$_x$. Such a multi-gap picture is supported by spin-lattice relaxation,\cite{Kawasaki2008} resistivity,\cite{Hunte2008} penetration depth,\cite{Martin2009} and point-contact spectroscopy\cite{Gonnelli2009} measurements. Although it has been widely assumed that the pairing is of spin-singlet s$\pm$ symmetry,\cite{Grafe2008,Mazin2008} with a sign change of the order parameter between the electron and hole pockets, questions regarding gap structure remain open after many years of research. Some experimental results\cite{Gonnelli2009,Oka2012} suggest a full gap, while others\cite{Martin2009,Sato2008} indicate a marked difference from an isotropic s-wave symmetry. More information is necessary to fully understand this issue.
     
Infrared spectroscopy is a well-known tool for elucidating the energy gap in conventional superconductors\cite{Glover1956} and in probing the electrodynamics in cuprates.\cite{Basov2005} It has been used by several groups to study pnictide superconductors, including LaFeAsO$_{1-x}$F$_x$. For this compound, experiments \cite{Dongi2008,Chen2008,Drechsler2008,Boris2009} have been predominantly on polycrystalline samples because single crystals are difficult to grow. Though infrared reflectance measurements have found the signature of a superconducting gap,\cite{Chen2008} transmission through a thin-film sample is expected to be more sensitive to the gap, perhaps best illustrated in BCS superconductors.\cite{Glover1956} Such data for LaFeAsO$_{1-x}$F$_x$ have not been reported in the literature. We report here the first infrared transmission spectroscopy results for LaFeAsO$_{1-x}$F$_x$ thin films and obtain consistent signatures of a full gap by both conventional and photo-induced time-resolved methods.

We measured the infrared transmission of LaFeAsO$_{1-x}$F$_x$ thin films in the superconducting and normal state. The $\sim$300~nm thick films were grown on a 1~mm thick single crystal LaAlO$_3$ substrate in a two-step process employing pulsed laser deposition.\cite{Haindl2013} $T_c$ was determined to be approximately 30~K from resistivity measurements. Two samples were studied, showing similar results. Here we present data on one sample. Infrared transmission was measured at Beamline U4IR of the National Synchrotron Light Source (NSLS, Brookhaven National Laboratory). The sample was mounted in a $^4$He Oxford cryostat and directly cooled in helium exchange gas. Infrared spectra were collected using a Bruker 66v spectrometer with a 1.5~K bolometric detector, spanning 20--120~cm$^{-1}$. Thick quartz cryostat windows prevented measurements at higher frequencies.

\begin{figure}[t]
\includegraphics[scale=1.2]{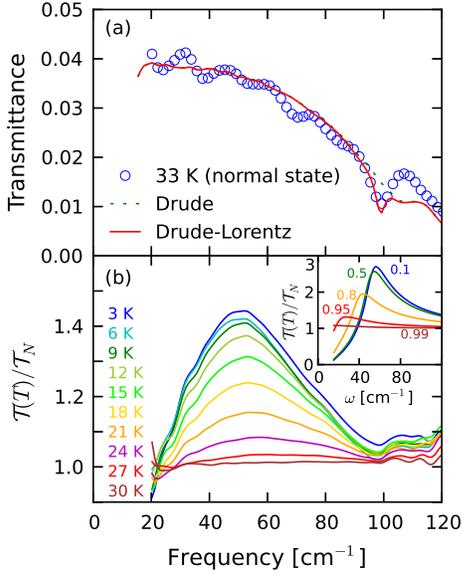}   
\caption{(a) The measured normal-state transmittance (circles) at 33~K, fitted to the Drude model (dashed line) and Drude-Lorentz model (solid line). (b) The temperature dependent transmission, normalized to the transmission at 33~K. The inset in (b) shows the temperature dependence of $\mathcal{T}(T)/\mathcal{T}_N$ had the superconductor been a BCS type with $2\Delta_0=50$~cm$^{-1}$. The numbers are values of $T/T_c$.} 
\label{FIG1}
\end{figure}

We first measured the transmittance of the sample in the normal state at 33~K (above $T_c$), shown in Fig.~\ref{FIG1}a. To analyze the transmittance of the film-on-substrate combination, we measured the transmittance and reflectance of a bare substrate. Both are almost independent of temperature between 4~K and 35~K in the 20--120~cm$^{-1}$ range (variation less than 0.1\%). The refractive index and extinction coefficient are extracted from these data using a standard procedure.\cite{Zhang1994} The sample transmittance closely follows the profile of the substrate transmittance, but its magnitude is much lower due to the absorption in the film. Using the substrate optical constants and the thin-film infrared transmission formula,\cite{Xi2010} the normal-state transmittance is fitted using an optical conductivity based on a combination of Drude and Lorentzian terms for describing both free carriers and low energy phonons,
\begin{equation}
\sigma_N(\omega) = \frac{\Omega_{p,N}^2/4\pi\gamma}{1-i\omega\gamma}+\sum_j\frac{\Omega_j^2\omega/4\pi i}{\omega_j^2-\omega^2-i\omega\gamma_j}.\label{eq1}
\end{equation}
Here $\sigma_N(\omega)$ is the normal-state optical conductivity at photon frequency $\omega$, with $\Omega_{p,N}$ and $\gamma$ the Drude plasma frequency and scattering rate, and $\Omega_j$, $\omega_j$, $\gamma_j$ denoting the plasma frequency, resonance frequency, linewidth of the $j$th Lorentzian oscillator. Due to the limited frequency range of the data, we can only assign a lower limit for the Drude scattering rate. In the analysis of the superconducting-state transmission (to be discussed below), we find that a $\gamma\ge 800$~cm$^{-1}$ is consistent with both the superconducting and normal state transmission results. A fit using $\gamma= 800$~cm$^{-1}$ yields a Drude plasma frequency of $\Omega_{p,N}=4595$~cm$^{-1}$.\cite{Note1} One Lorentzian term fits the phonon mode, yielding the phonon plasma frequency $\Omega_{ph}=200$~cm$^{-1}$, resonance frequency $\omega_{ph}=99$~cm$^{-1}$, and linewidth $\gamma_{ph}=5$~cm$^{-1}$. Fits with and without the Lorentzian term are compared in Fig.~\ref{FIG1}a. Such a low frequency phonon, previously observed in polycrystalline materials,\cite{Dongi2008,Drechsler2008} have been shown to originate from vibrations involving principally La, Fe and As atoms.\cite{Boeri2008} 

\begin{figure}[t]
\includegraphics[scale=1.1]{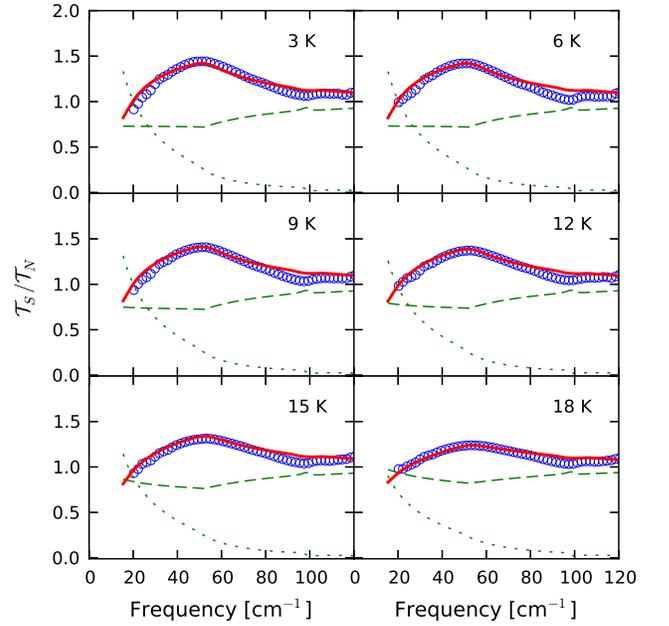}                
\caption{The circles are measured superconducting-to-normal transmission ratios. The solid line in Panel a is a two-component fit. The solid lines in Panels b--f are two-component calculations using the fitting parameters and the BCS temperature dependence of the gap. The dashed and dotted lines in each panel are the real and imaginary parts of the superconducting optical conductivity at the corresponding temperature, normalized to $\sigma_N(\omega)$.} 
\label{FIG2}
\end{figure}   

We next measured the transmission of the sample in the superconducting state from 3~K to 30~K, normalized to the 33~K spectrum, with the results shown in Fig.~\ref{FIG1}b. The ratio shows a peak similar to what is observed for a conventional s-wave superconductor; the peak position would then suggest an optical gap $2\Delta_0\sim$50~cm$^{-1}$. As the temperature is increased, the ratio approaches unity above 27~K, consistent with $T_c=30$~K. However, the peak position does not change significantly with temperature, and the peak amplitude at $T\ll T_c$ is much lower than that for a conventional s-wave superconductor of this optical thickness. For comparison, we calculated $\mathcal{T}(T)/\mathcal{T}_N$ for a dirty-limit weak-coupling BCS superconducting film of the same sheet resistance, assuming a zero-temperature gap $2\Delta_0=50$~cm$^{-1}$. The results for $T/T_c=0.1$, 0.5, 0.8, 0.95, and 0.99 are shown in the inset of Fig.~\ref{FIG1}b. The different temperature dependence from LaFeAsO$_{1-x}$F$_{x}$ is apparent. In light of the band structure's separate electron and hole pockets, we considered a two-component model based on a non-superconducting Drude term and a BCS Mattis-Bardeen superconducting term, and found good agreement with the experimental data. Such a model was first proposed by Lobo \textit{et al.} when discussing a single-crystal Ba(Fe$_{1-x}$Co$_x$)$_2$As$_2$ sample.\cite{Lobo2010} Slightly modified from their form, we use
\begin{align}
\sigma(\omega) & = \frac{\Omega_{p,D}^2/4\pi\gamma}{1-i\omega\gamma}+\sigma_{\mathrm{BCS}}(\omega,\Omega_{p,S},\gamma,\Delta_0,T)\nonumber\\
&\quad +\frac{\Omega_{ph}^2\omega/4\pi i}{\omega_{ph}^2-\omega^2-i\omega\gamma_{ph}}.\label{eq2}
\end{align}
Here the parameters in the Drude and Lorentzian terms have the same meaning as those in Eq.~\eqref{eq1}. We constrain the two plasma frequencies $\Omega_{p,S}$ and $\Omega_{p,D}$ to agree with the normal state (i.e. $\Omega_{p,S}^2+\Omega_{p,D}^2=\Omega_{p,N}^2$). By fitting the $\mathcal{T}_S/\mathcal{T}_N$ at 3~K (Panel a in Fig.~\ref{FIG2}), we find that the same scattering rate $\gamma\ge 800$~cm$^{-1}$ found for the normal state is obtained for both Drude and BCS terms in the superconducting state. Setting $\gamma=800$~cm$^{-1}$ yields the Drude plasma frequency $\Omega_{p,D}=3844$~cm$^{-1}$. We use the formalism by Zimmerman \textit{et al}. for the BCS superconducting term,\cite{Zimmermann1991} in which $\Omega_{p,S}=2517$~cm$^{-1}$. The fit yields a zero-temperature energy gap $2\Delta_0=52$~cm$^{-1}$, consistent with infrared reflectance measurement.\cite{Chen2008} For simplicity, the Lorentzian term is assumed to be the same as that in the normal state. Using the above fitting parameters and assuming a weak-coupling BCS temperature dependence for the gap, we calculated 
$\mathcal{T}_S/\mathcal{T}_N$ from 6~K to 18~K, shown as solid lines in Panels b--f in Fig.~\ref{FIG2}. The results are consistent with the experimental data. The dashed and dotted lines in Fig.~\ref{FIG2} are the real and imaginary parts of the fitted or calculated optical conductivity at the corresponding temperatures, both normalized to the normal-state conductivity $\sigma_N(\omega)$. The real part shows significant residual absorption that lowers the superconducting-state transmission to yield a $\mathcal{T}_S/\mathcal{T}_N$ peak amplitude below the Mattis-Bardeen prediction. The origin of this non-gapped conductivity component may be similar to that discussed by Lobo \textit{et al}.\cite{Lobo2010} (and references therein). If the residual absorption is due to the electron pockets, then the full gap would exist in the hole pockets. 
  
\begin{figure}[t]
\includegraphics[scale=1.2]{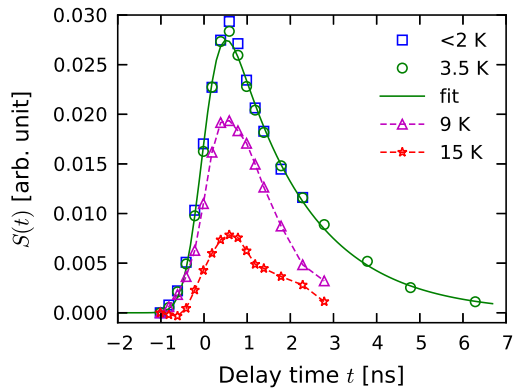}                
\caption{Time-resolved photo-induced transmission at various temperatures. Squares are data acquired with the sample immersed in superfluid $^4$He ($T<2$~K). Data at other temperatures were collected with the sample in helium exchange gas. The solid line is a fit as described in the text.} 
\label{FIG3}
\end{figure}

Further evidence of a full gap is provided by a laser-pump infrared-probe experiment. The sample was mounted in the cryostat in the same transmission configuration as described above. Near-infrared pulses from a mode-locked Ti-sapphire laser were delivered over fiber optical cable to the sample for photoexcitation. The far-infrared portion of the synchrotron radiation, the same as that used in the conventional transmission spectroscopy, was used to probe the change of the film's transmission after photoexcitation. Because electrons move in bunches in the synchrotron storage ring, they produce synchrotron radiation in pulses. The pulse width determines the time resolution to be $\sim$300~ps. Different from most all-optical pump-probe studies of pnictides in the literature,\cite{Mertelj2009,Mansart2010,Chia2010,Torchinsky2010} where the probe is near-infrared laser pulses, our technique is expected to be more sensitive to the photo-induced change on the energy scale of the superconducting gap. The relative delay between the laser and synchrotron pulses was controlled from a pulse generator, so that the relaxation after photoexcitation can be studied in real time. Details of the setup were reported elsewhere.\cite{Xi2013} A laser pulse fluence of 1.6~nJ/cm$^2$ was used, corresponding to an average intensity of 85~mw/cm$^2$ and well below a level where significant heating occurs. The photo-induced transmission signal $S(t)$, which is assumed to be proportional to the excess quasiparticle density, is shown in Fig.~\ref{FIG3} for a few temperatures. The film thickness and sample size limited the signal-to-noise ratio in the measurement, but the decay transients were reproducible. To further rule out thermal effects, we ran one measurement with the sample fully immersed in superfluid $^4$He at temperature below 2~K, so that heating could be minimized. The decay behavior was confirmed. (Compare squares and circles in Fig.~\ref{FIG3}.) 

The photo-induced decay indicates a nanosecond effective quasiparticle recombination time, consistent with a full gap (absence of nodes). Each trace shown in Fig.~\ref{FIG3} is a convolution of the sample intrinsic relaxation signal and the synchrotron probe pulse. Assuming an exponential decay for the former and a gaussian function for the latter, we extracted a lifetime of 1.8$\pm$0.2~ns for the excess quasiparticles at 3.5~K. (See the solid line in FIG.~\ref{FIG3} for the fit.) Such a time scale is comparable to the effective quasiparticle lifetimes in BCS superconductors. In these superconductors, the effective quasiparticle recombination is typically dominated by a phonon bottleneck effect, first explained by Rothwarf and Taylor.\cite{Rothwarf1967} The phonons emitted in a quasiparticle recombination process typically re-break Cooper pairs before decaying to energy lower than 2$\Delta$ or leaving the sample. Such an effect increases the quasiparticle lifetime dramatically to values approaching 1~ns and greater.\cite{Carr2000,Demsar2003,Lobo2005} Such a long effective relaxation time has not been reported in pump-probe experiments on pnictide superconductors. In a study\cite{Mertelj2009} of SmFeAsO$_{0.8}$F$_{0.2}$, a 5~ps relaxation time was ascribed to superconductivity and a sub-ps relaxation was related to the pseudogap. Relaxation times of tens of ps or less were consistently reported in optical pump-probe experiments of both electron and hole doped BaFe$_2$As$_2$.\cite{Mansart2010,Chia2010,Torchinsky2010} One of them\cite{Torchinsky2010} also reported a slow decay of hundreds of ps and attributed that to the bottleneck effect in one of the fully-gapped hole bands. We argue that the nanosecond decay observed in our sample is of the same origin: existence of a full gap combined with a bottleneck effect. A nodal gap provides more electronic states to be involved in both quasiparticle and phonon scattering, leading to a much faster relaxation process. This argument does not rule out the existence of a nodal gap in another band. As long as the coupling between such a band and the band with the full gap is weak, the bottleneck effect would still be present.

\begin{figure}[t]
\includegraphics[scale=1.2]{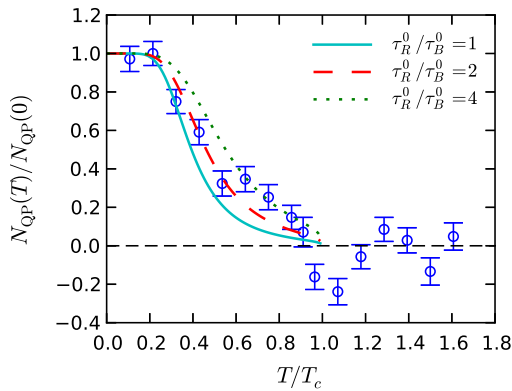}                
\caption{Temperature dependence of the normalized maximum gap-edge excess quasiparticle density. The error bars indicate the standard deviation of the signal in the measurement. The lines are calculations using Eq.~\eqref{eq3}.} 
\label{FIG4}
\end{figure}  

To gain more information about the quasiparticle relaxation, we measured the peak photo-induced transmission signal $S_{\mathrm{max}}$ at a fixed delay setting as a function of temperature. The data are plotted in Fig.~\ref{FIG4}, normalized to the peak value at low temperature. In the figure we label this temperature-dependent $S_{\mathrm{max}}(T)$ as the maximum gap-edge excess quasiparticle density $N_{\mathrm{QP}}(T)$, because to the first-order approximation $S_{\mathrm{max}}\propto N_{\mathrm{QP}} $.\cite{Carr2000,Lobo2005} $N_{\mathrm{QP}}(T)$ decreases as the temperature increases towards $T_c$. Above $T_c$, the origin of the negative signal is not clear. It could be due to fluctuation effects close to $T_c$, but given the low signal-to-noise ratio in the measurement, it could be a thermal artifact. Measurements with much higher signal-to-noise ratio will be useful to clarify this. We focus on the data below $T_c$ and adopt an analysis that was successfully used to study BCS superconductors.\cite{Nicol2003,Lobo2005} When the laser energy is absorbed by the superconductor, it first breaks Cooper pairs and creates high-energy quasiparticles. These quasiparticles scatter among themselves and with phonons, equilibrating to the gap edge typically within picoseconds. Due to the $\sim$300~ps time resolution in our measurement, we only probe the relaxation after the quasiparticles and the phonons have equilibrated to the gap edge. When a phonon bottleneck is present, the trapped energy from the laser pulse is shared between the populations of excess quasiparticles and phonons. In this quasi-equilibrium condition, the temperature-dependent fraction of energy in the quasiparticles obeys\cite{Carr2000,Lobo2005}
\begin{equation}
\frac{N_{\mathrm{QP}}(T)}{N_{\mathrm{QP}}(0)}=\frac{\Delta_0}{\Delta(T)}\frac{1}{1+2\tau_B(T)/\tau_R(T)}.\label{eq3}
\end{equation}
Here $\tau_R$ and $\tau_B$ are the near-equilibrium intrinsic quasiparticle recombination lifetime and the phonon pair-breaking time, respectively, as discussed by Kaplan \textit{et al}.\cite{Kaplan1976} They involve the quasiparticle and phonon density of states, the coherence factor, the gap temperature dependence, as well as material dependent coefficients $\tau_R^0$ and $\tau_B^0$. Assuming a BCS dependence for the various quantities, we calculated $N_{\mathrm{QP}}(T)/N_{\mathrm{QP}}(0)$ according to Eq.~\eqref{eq3}, with $\tau_R^0/\tau_B^0$ as the only adjustable parameter. The results are shown in Fig.~\ref{FIG4}, consistent with the trend shown by the data up to $T_c$. The same type of measurement shows a similar trend in BCS superconductors,\cite{Lobo2005} but in the cuprates\cite{Kabanov1999} the flat portion extends to a much higher reduced temperature $T/T_c$, followed by a sudden drop. If the above analysis is valid, it confirms the full gap and phonon bottleneck proposed in the previous paragraph. It also implies a BCS temperature dependence to the gap.

In conclusion, we performed conventional and time-resolved infrared spectroscopy on LaFeAsO$_{1-x}$F$_x$ thin films. The former suggests a full gap and residual absorption. The latter points to a bottleneck picture in the presence of a full gap. The two experiments provide convergent evidence of a full gap in LaFeAsO$_{1-x}$F$_x$. Weak-coupling BCS theory gives a good description of our data, but it is not totally validated for this compound.\cite{Boeri2008} However, the nanosecond quasiparticle relaxation time observed here is independent of such analysis and strongly supports our conclusion. More information is required to assign this gap to a specific band. A theoretical proposal demonstrated that the existence of a nodeless gap in an extended s-wave system is possible if significant disorder is present.\cite{Mishra2009,Carbotte2010} If samples with controlled levels of disorder are available, a time-resolved study as reported here will be a sensitive test to the theory.

We thank Qiang Li for providing a bare substrate. This work was supported by the U.S. Department of Energy through contract DE-ACO2-98CH10886. S. H. and M. K. acknowledge financial funding by the German Research Foundation under project HA5934/1-1.

\end{document}